\documentstyle[ltwol]{article}

\def\Journal#1#2#3#4{{#1} {\bf #2}, #3 (#4)}

\def\NPB{{\em Nucl. Phys.} B}
\def\PLB{{\em Phys. Lett.}  B}
\def\PRL{\em Phys. Rev. Lett.}
\def\PRD{{\em Phys. Rev.} D}

\def\be{\begin{equation}}
\def\ee{\end{equation}}
\def\bea{\begin{eqnarray}}
\def\eea{\end{eqnarray}}

\input psfig.sty

\begin{document}

\title{FIXED POINT STRUCTURE OF PAD\'E-SUMMATION APPROXIMATIONS
TO THE QCD $\beta$-FUNCTION}

\author{V. Elias$^*$, F. Chishtie}
\address{Department of Applied Mathematics, The University of Western
Ontario, London, Ontario  N6A 5B7, CANADA\\$^*$E-mail: zohar@apmaths.uwo.ca}   

\author{T.G. Steele}
\address{Department of Physics and Engineering Physics, University of
Saskatchewan, Saskatoon, Saskatchewan  S7N 5C6\\
E-mail: steelet@sask.usask.ca}

\twocolumn[\maketitle\abstracts{Pad\'e-improvement of four-loop
$\beta$-functions in massive $\phi^4$ scalar field theory is shown to
predict the known five-loop contribution with astonishing (0.2\%)
accuracy, supporting the applicability of Pad\'e-summations for
approximating all-orders $\overline{\rm MS}$ QCD $\beta$-functions, as
suggested by Ellis, Karliner, and Samuel.  Surprisingly, the most
general set of $[2|2]$ approximants consistent with known two-, three-,
and four-loop contributions to the QCD $\beta$-function with up to
six flavours fail to exhibit any zeros that could be interpreted as
positive infrared fixed points, {\it regardless} of the unknown
five-loop term.  When they occur, positive zeros of such $[2|2]$
approximants are preceded by singularities, leading to a
double-valued $\beta$-function that is decoupled entirely 
from the infrared region,
similar to the $\beta$-function of SUSY gluodynamics.
}]

\begin{figure}[htb]
\psfig{figure=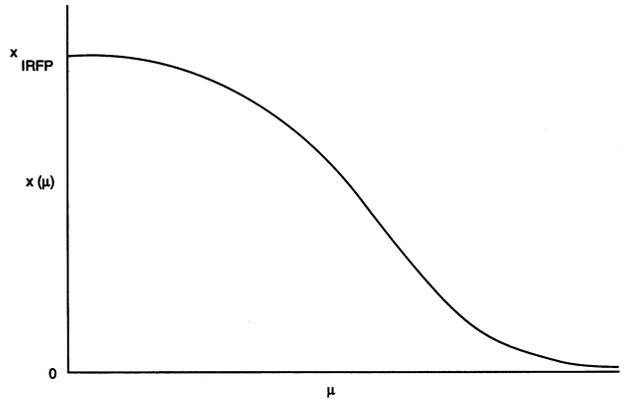,width=3.5in}
\caption{Mattingly-Stevenson scenario}
\label{fig1}
\end{figure}

Higher order terms of the QCD $\overline{\rm MS}$ $\beta$-function
\renewcommand{\theequation}{1\alph{equation}}
\setcounter{equation}{0}
\begin{equation}
\mu^2 \frac{dx}{d\mu^2} \equiv \beta(x),
\label{1a}
\end{equation}
\begin{equation}
\beta(x) = - \sum_{i=0}^\infty \beta_i x^{i+2},
\label{1b}
\end{equation}
$x \equiv \alpha_s (\mu) / \pi$ are, upon truncation, known to permit
the occurrence of fixed points other than the ultraviolet fixed point
at $x = 0$; e.g. the positive infrared fixed point (IRFP) which
occurs for $9 \leq n_f \leq 16$ when the series for $\beta(x)$ in (1)
is truncated after two terms [$\beta_0 = (11 - 2n_f / 3) / 4$;
$\beta_1 = (102 - 38n_f / 3) / 16$; $x_{IRFP} = -\beta_0 / \beta_1$]. 
However, the fixed points arising from such truncation are likely to
be spurious, as the candidate-value for $x_{IRFP}$ is sufficiently
large for the highest-order term in the series $\beta(x_{IRFP})$ to
be comparable in magnitude to lower terms [e.g.$|\beta_1 x^3| =
|\beta_0 x^2|$].  In a recent paper, \cite{je} Ellis, Karliner and
Samuel predicted the coefficient $\beta_3$ via Pad\'e approximant
methods, and claimed that $\beta_{0-2}$ and their prediction for
$\beta_3$ yield a Pad\'e summation of the $\beta$-function with a
nonzero IRFP consistent with an earlier prediction by Mattingly and
Stevenson.\cite{acm}  This Mattingly-Stevenson scenario leads to the
freezing-out of the coupling to a constant value in the infrared
region, as shown schematically in Fig. \ref{fig1}.

Pad\'e summation of the $\overline{MS}$ $\beta$-function identifies
the infinte series
\renewcommand{\theequation}{\arabic{equation}}
\setcounter{equation}{1}
\begin{equation}
\beta(x) = -\beta_0 x^2 (1+R_1 x + R_2 x^2 + R_3 x^2 + ...)
\label{2}
\end{equation}
$(R_i \equiv \beta_i / \beta_0)$ with a Pad\'e
approximant which incorporates the known UV asymptotics
of the $\beta$-function,
\renewcommand{\theequation}{3\alph{equation}}
\setcounter{equation}{0}
\begin{equation}
\beta(x) \rightarrow \beta_{[N|M]} (x) \equiv -\beta_0 x^2 S_{[N|M]}
(x),
\label{3a}
\end{equation}
whose Maclaurin expansion reproduces the known terms in the infinite
series (2):
\begin{eqnarray}
S_{[N|M]} (x) & = & \frac{1+a_1 x + \ldots + a_N x^N}{1 + b_1 x + \ldots + b_M
x^M} \nonumber \\
& = & 1 + R_1 x + R_2 x^2 + R_3 x^3 + \ldots\; \; .
\label{3b}
\end{eqnarray}
An IRFP of the $\beta$-function would, in this approximation,
necessarily be identified with a positive zero of $\beta_{[N|M]}$; 
i.e. a positive zero $(x_{num})$ of $1+a_1 x + ... + a_N x^N$, the
numerator of $S_{[N|M]}$, {\it provided $S_{[N|M]}$ remains positive
for $0 \leq x \leq x_{num}$}.  This latter requirement precludes the
existence of a positive zero $(x_{den})$ of the denominator $1+b_1 x
+ ... + b_M x^M$ that falls in the interval $0 \leq x \leq x_{num}$.

\begin{figure}[htb]
\psfig{figure=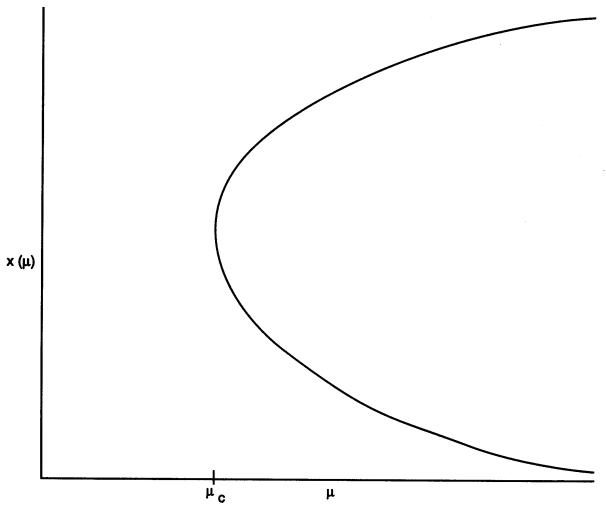,width=3.5in}
\caption{The Kogan-Shifman scenario $\left[x_{den}=x(\mu_c)\right]$}
\label{fig2}
\end{figure}

One cannot automatically dismiss the possibility of such a denominator
zero occuring within the true QCD $\beta$-function.  The
$\beta$-function of SU(N) SUSY gluodynamics is known {\it exactly} if
no matter fields are present,\cite{nsv} and exhibits precisely
such a zero:
\renewcommand{\theequation}{\arabic{equation}}
\setcounter{equation}{3}
\begin{equation}
\beta(x) = -\frac{3Nx^2}{4} \left[ \frac{1}{1 - N x / 2} \right].
\label{4}
\end{equation}
The $\beta$-function (4) has been discussed further by Kogan and
Shifman.\cite{ks}  If (4) is incorporated into (1a), the resulting
Kogan-Shifman scenario (Fig. \ref{fig2}) for $x(\mu)$ is indicative of both a
strong phase in the ultraviolet region (the upper branch of Fig. 2)
as well as the existence of an infrared cut off $(\mu_c)$ on the
domain of $x(\mu)$ that renders the infrared region $\mu < \mu_c$
inaccessible. \footnote{We are assuming $\alpha_s$ to be real.  The
possibility of $\alpha_s$ being complex for $\mu < \mu_c$ is
addressed in ref. 4.} 

Without use of Pad\'e summation methods, as described above, the
known terms of the infinite series representation  (2) for the $\overline{\rm MS}$ QCD
$\beta$-function offer little information as to whether the
Mattingly-Stevenson (Fig. 1) or Kogan-Shifman (Fig. 2) scenario is
appropriate for the evolution of the strong coupling.  However, there
is reason to believe that Pad\'e summation representations of the
$\beta$-function [eq. (3a)] may indeed be appropriate for quantum
field theoretical calculations.  Ellis, Gardi, Karliner, and Samuel
\cite{je,sek} have argued that Pad\'e-summations (3) converge to
their perturbative series (2) as $N$ and $M$ increase for any such
series dominated by a finite set of renormalon poles, consistent with
the following asymptotic error formula for the difference between
$R_{N+M+1}$ in (2) and the value $R_{N+M+1}^{\rm Pad\acute{e}}$ predicted
via use of the $[N|M]$ approximant in (3b): \cite{je,ejj}
\begin{eqnarray}
\delta_{N+M+1} & \equiv & \frac{R_{N+M+1}^{\rm Pad\acute{e}} -
R_{N+M+1}}{R_{N+M+1}} \nonumber \\ 
& = & -\frac{M!A^M}{[N+M(1+a) + b]^M}.
\label{5}
\end{eqnarray}
In (5), $\{a,b,A\}$ are constants to be determined.

To demonstrate the utility of this asymptotic error formula, consider
the known $\beta$-function for massive $\phi^4$ scalar-field
theory:\cite{kns}
\begin{equation}
{\cal{L}} = \frac{1}{2} (\partial_\mu \phi) (\partial^\mu \phi) +
\frac{1}{2} m^2 \phi^2 + g \left( \frac{16\pi^2}{4!} \right)
(\phi^2)^2,
\label{6}
\end{equation}
\begin{eqnarray}
\beta(g) & = & 1.5 g^2 \left\{ 1 - \frac{17}{9}g + 10.8499g^2 -
90.5353g^3 \right. \nonumber \\
& + & \left. R_4g^4+ ... \right\}.
\label{7}
\end{eqnarray}
Using the first two terms of (7) to generate a $[0|1]$ approximant,
as in (3), one would predict $R_2^{Pad\acute{e}} =  (-17/9)^2$, in
which case we see from (5) that
\begin{equation}
\delta_2 = \frac{(-17/9)^2 - 10.8499}{10.8499} = \frac{-A}{1+(a+b)}.
\label{8}
\end{equation}
Using the first three terms of (7) to generate a $[1|1]$ approximant,
one would predict $R_4^{\rm Pad\acute{e}} = (10.8499)^2 / (-17/9)$, in
which case
\begin{eqnarray}
\delta_3 & = & \frac{[(10.8499)^2 / (-17/9)] -
(-90.5353)}{(-90.5353)}\nonumber \\
& = & \frac{-A}{2+(a+b)}.
\label{9}
\end{eqnarray}
Equations (8) and (9) are two
equations for the unknown constants $A$ and $(a+b)$, with solutions
\begin{equation}
A = \left[ 1 / \delta_2 - 1 / \delta_3\right]^{-1}, (a+b) = (\delta_2
- 2\delta_3) / (\delta_3 - \delta_2).
\label{10}
\end{equation}

We can substitute (10) into (5) to determine $R_4$.  The first three
terms in the series (7) generate a $[2|1]$ approximant whose
Maclaurin expansion (3b) predicts $R_4^{\rm Pad\acute{e}} =
(-90.5353)^2 / (10.8499)$. Upon substitution of $A$, $(a+b)$, and
$R_4^{\rm Pad\acute{e}}$ into (5) we find that this asymptotic error
formula {\it predicts} that 
\begin{eqnarray}
R_4 & = & R_4^{\rm Pad\acute{e}}/ (1 + \delta_4) 
\label{11}
\\
& = & R_4^{\rm Pad\acute{e}} [3+(a+b)] / [3 + (a+b)\!-\! A] = 947.8
\nonumber 
\end{eqnarray}
The value of $R_4$ in (7) has been explicitly calculated \cite{kns}
to be 949.5, in very close agreement with (11).  The series of steps
leading from (5) to (11), a methodological recipe first presented in
ref. 6, has also been applied \cite{je,vsc}to $N$-component scalar
field theory,
for which the Lagrangian (6) is modified such that $\phi\rightarrow\phi_a$,
$\phi^2\rightarrow \sum_{a=1}^{N} \phi^a \phi^a$, $N =
\{ 2,3,4 \}$.  Agreement with calculated values of $R_4$  $(R_4\equiv
\beta_4 / \beta_0)$ remains within 3.5\% for $N \leq 4$.\cite{vsc}

This startling agreement suggests that Pad\'e methodology may also
be applicable to the QCD $\beta$-function, particularly in the $n_f =
0$ gluodynamic limit where 
\begin{enumerate}
\item such methods are expected to be most
accurate,\cite{ejj}  
\item  comparison with the Kogan-Shifman scenario
for SUSY gluodynmamics is most relevant.  
\end{enumerate}
For $n_f = 0$, the
4-loop
$\overline{\rm MS}$ QCD $\beta$-function, as defined by (1a), is
given by \cite{rvl}
\begin{eqnarray}
\beta(x) & = & -\frac{11}{4} x^2 [1+2.31818 + 8.11648x^2 \nonumber \\
& + & 41.5383x^3 + \sum_{k=4}^\infty R_k x^k].
\label{12}
\end{eqnarray}
The coefficients $R_k$ 
are presently not known for $k \ge 4$.  
The
first three terms in the series (12) are sufficient in themselves to
determine the Pad\'e approximants $S_{[1|2]}$ and $S_{[2|1]}$, as
defined in (3).  These approximants are
\begin{equation}
\beta_{[2|1]} (x) = -\frac{11}{4}x^2 \left[ \frac{1-2.7996x -
3.7475x^2}{1-5.1178x} \right],
\label{13}
\end{equation}
\begin{equation}
\beta_{[1|2]} (x) = -\frac{11}{4}x^2 \left[
\frac{1-5.9672x}{1-8.2854x + 11.091x^2} \right].
\label{14}
\end{equation}
In both (13) and (14), the (first) positive denominator zero {\it
precedes} the positive numerator zero:
for (13), $x_{num} = 0.264 > x_{den} = 0.195$;  for (14), $x_{num} =
0.168 > x_{den} = 0.151$.  Consequently, $x_{num}$ cannot be
identified with the Mattingly-Stevenson IRFP in either case, as this
zero is separated from the small x-region by a singularity past which
the $\beta$-function switches sign.  Indeed the ordering $0 < x_{den}
< x_{num}$ is suggestive of a Kogan-Shifman scenario in which
$x_{num}$, if taken seriously, is an {\it ultraviolet} fixed point
(UVFP) characterizing the strong phase [{\it i.e.}, the upper branch
of Fig 2].

We can apply the asymptotic error formula (5) to the series (12) in
precisely the same way we applied it to (7).  We then obtain an
estimate $R_4 = 302.2$, analogous to (11).  Using this value of $R_4$
in conjunction with the known terms of (12), it is possible to obtain
a $[2|2]$-approximant $\beta$-function
\begin{equation}
\beta_{[2|2]}(x) = -\frac{11}{4} x^2 \left[ \frac{1-9.6296x +
4.3327x^2}{1-11.9477x + 23.913x^2} \right].
\label{15}
\end{equation}
The first positive numerator zero $x_{num} = 0.1092$ is again larger
than the first positive denominator zero $x_{den} = 0.1063$,
precluding the identification of $x_{num}$ as the IRFP of the
Mattingly-Stevenson scenario (Fig. 1). Instead, the $\beta$-function
(15) is consistent with the Kogan-Shifman scenario of Fig. 2, with
$x_{num}$ again identified as a nonzero UVFP for the strong phase.

\begin{figure}[htb]
\psfig{figure=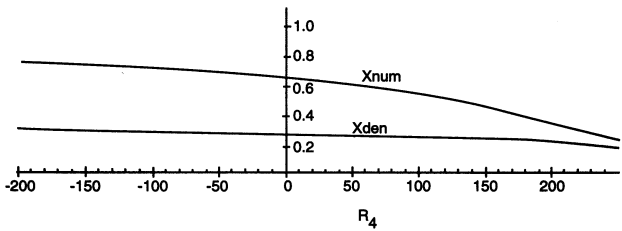,width=3.5in}
\caption{Dependence of the first positive numerator zero ($x_{num}$)
and denominator zero ($x_{den}$) of (16) on $R_4$, the (presently unknown)
five-loop coefficient of the $n_f=0$ $\beta$-function.
}
\label{fig3}
\end{figure}

Curiously, the ordering $0 < x_{den} < x_{num}$ characterizes
$[2|2]$-approximant $\beta$-functions {\it even if $R_4$ is allowed
to be arbitrary}.  The most general such $\beta$-function that
reproduces the first four terms of (12) [the first three being known]
is
\begin{eqnarray}
& &\beta_{[2|2]} (x) = -\frac{11}{4} x^2 \times 
\label{16}\\
& &\left[
\frac{1+(13.403-0.076215R_4)x-(22.915-0.090166 R_4)x^2}{1+(11.084-0.076215
R_4)x - (56.727-0.26685 R_4)x^2} \right].
\nonumber
\end{eqnarray}
It is easy to verify the first positive numerator zero of (16) is
always larger than the first positive denominator zero
[Fig. \ref{fig3}], although
these zeros become asymptotically close as $R_4 \rightarrow +\infty$. 
Thus, we see that the first positive zero of {\it any} $[2|2]$ Pad\'e
approximant whose Maclaurin expansion reproduces the known terms of
eq. (12) {\it cannot} be identified as an IRFP, nor is such
Pad\'e-summation indicative of a Fig. 1 scenario for the
$\overline{\rm MS}$ $n_f = 0$ $\beta$-function.

Remarkably, the same set of conclusions can be drawn for the
physically interesting case of three light flavours.  When $n_f = 3$,
the 4-loop $\overline{\rm MS}$ QCD $\beta$-function is given
by \cite{rvl}
\begin{eqnarray}
\beta(x) & = & -\frac{9x^2}{4} \left[ 1 + (16/9)x + 4.471065x^2 \right.
\nonumber\\
& + & \left. 20.99027x^3 + \sum_{k=4}^\infty R_k x^k \right],
\label{17}
\end{eqnarray}
with $R_k$ not presently known for $k \geq 4$.  The known terms in
(17) determine $[2|1]$ and $[1|2]$ Pad\'e-summation representations
of the $n_f = 3$ $\beta$-function,
\begin{equation}
\beta_{[2|1]} (x) = - \frac{9x^2}{4} \left[ \frac{1 - 2.91691x 
- 3.87504x^2}{1 - 4.69468x} \right],
\label{18}
\end{equation}
\begin{equation}
\beta_{[1|2]}(x) = -\frac{9x^2}{4} \left[
\frac{1-8.17337x}{1-9.95115x+13.2199x^2} \right].
\label{19}
\end{equation}
The positive zero of $\beta_{[2|1]}(x)$ $(x=0.2559)$ occurs after the
pole at $x = 0.2130$; the positive zero of $\beta_{[2|1]} (x)$ at $x
= 0.1223$ similarly occurs after a pole at $x = 0.1194$.  The most
general ${[2|2]}$ approximant consistent with (17) is
\begin{eqnarray}
& & \beta_{[2|2]}(x) = -\frac{9x^2}{4} \times \\
& & \left[ \frac{1+(7.1945-0.10261R_4)x - (11.329-0.075643R_4)x^2}{
1+(5.4168-0.10261R_4)x - (25.430-0.25806R_4)x^2} \right] . \nonumber
\label{20}
\end{eqnarray}
The Maclaurin expansion of (20) reproduces the series in (17),
including its (unknown) $R_4x^4$ term.  As was the case in (16), the
first positive zero of the denominator of (20) is always seen to
precede the first positive zero of the numerator, regardless of
$R_4$.  Thus the $[1|2]$, $[2|1]$ and most general possible
$[2|2]$-approximant representations of the $n_f = 3$ $\overline{\rm
MS}$ $\beta$-function uphold the ordering $0 < x_{den} < x_{num}$, an
ordering that precludes
the identification of $x_{num}$ with the IRFP of the 
Mattingly-Stevenson scenario.  Moreover, $[2|2]$-approximant
$\beta$-functions for arbitrary $R_4$ have been constructed
\cite{vsc} analogous to (16) and (20) for $n_f = \{4,5,6 \}$, and for each
of these, the $0 < x_{den} < x_{num}$ ordering persists regardless of
$R_4$.  A range for $R_4$ for which an ordering compatible with Fig. 1 
($0 < x_{num} < x_{den}$) is possible does not occur until $n_f = 7$.

As noted above, the ordering $0 < x_{den} < x_{num}$ suggests the
occurrence of a double-valued QCD coupling constant, as is the case in
SUSY gluodynamics (Fig. 2). Such a scenario is seen to decouple the
infrared region $\mu < \mu_c$ from the domain of $\alpha_s$, provided
$\alpha_s$ is understood to be real. Such a scenario is also indicative of a
strong phase at short distances \cite{ks} with possible implications
for dynamical electroweak symmetry breaking, suggesting that QCD may
even furnish its own ``technicolour.''

\section*{Acknowledgements}
VE and TGS are grateful for research support from NSERC, the Natural Sciences and Engineering
Research Council of Canada.

\end{document}